# Carbon doping of GaN: Proof of the formation of electrically active tri-carbon defects

I. Gamov,[1,a] E. Richter,[2] M. Weyers,[2] G. Gärtner,[3] and K. Irmscher[1,a]

[1]*Leibniz-Institut für Kristallzüchtung, Max-Born-Str. 2, 12489 Berlin, Germany*

[2]*Ferdinand-Braun-Institut, Leibniz-Institut für Höchstfrequenztechnik, Gustav-Kirchhoff-Str. 4, 12489 Berlin, Germany*

[3]*Institute of Experimental Physics, TU Bergakademie Freiberg, 09599 Freiberg, Germany*

ABSTRACT

Carbon doping is used to obtain semi-insulating GaN crystals. If the carbon doping concentration exceeds $5 \times 10^{17}$ cm$^{-3}$, the carbon atoms increasingly form triatomic clusters. The tri-carbon defect structure is unambiguously proven by the isotope effect on the defects' local vibrational modes (LVMs) originally found in samples containing carbon of natural isotopic composition (~99% $^{12}$C, ~1% $^{13}$C) at 1679 cm$^{-1}$ and 1718 cm$^{-1}$. Number, spectral positions, and intensities of the LVMs for samples enriched with the $^{13}$C isotope (~99% and ~50%) are consistently interpreted on the basis of the harmonic oscillator model taking into account the probability of possible isotope combinations. Including the polarization dependence of the LVM absorption, we show that the tri-carbon defects form a triatomic molecule-like structure in two crystallographically different configurations: a basal configuration with the carbon bonds near the basal plane and an axial configuration with one of the carbon bonds along the *c*-axis. Finally, the disappearance of the LVMs under additional below-bandgap illumination is interpreted as defect recharging, i.e. the tri-carbon defects possess at least one charge state transition level within the bandgap and contribute to optical absorption as well as to the electrical charge balance.

[a]Authors to whom correspondence should be addressed: Ivan.Gamov@ikz-berlin.de and Klaus.Irmscher@ikz-berlin.de

I. INTRODUCTION

Carbon doping is used in GaN-based high-power device technology to grow buffer layers by metalorganic vapor phase epitaxy for the isolation of device regions from parasitic conductive channels,[1,2,3,4] and in the growth of bulk crystals by hydride vapor phase epitaxy (HVPE) for the preparation of semi-insulating substrates.[5,6,7] Since in nominally undoped GaN crystals *n*-type conductivity prevails due to the presence of residual donor impurities such as oxygen or silicon, carbon doping must result in the formation of compensating acceptors.[8,9] Currently, a single carbon atom substituting for a nitrogen host atom ($C_N$) is considered the prevailing defect responsible for the compensation as $C_N$ possesses a deep acceptor level at ~1 eV above the valence band edge and has low formation energy compared to other carbon-related defects.[10–13,14,15,16,17] In addition, $C_{Ga}$ (carbon on gallium site), $C_i$ (interstitial carbon), and various carbon-related complexes with intrinsic defects or other impurities such as hydrogen, oxygen or silicon have been discussed, and due to their action as donors or (partly passivated) acceptors, they may be involved in the compensation process.[10–12,18] Among the defect complexes consisting of multiple carbon atoms only for carbon pairs theoretical predictions are available which report high to moderate pair formation energies making substantial defect concentrations under thermodynamical equilibrium conditions less probable.[11,13] However, the possibility of enhanced formation of defect complexes at the crystal's growing surface, and its subsequent overgrowth that might be accompanied by the freeze out of the defects in high non-equilibrium concentrations is usually not considered. In AlN crystals, where such multiple-carbon defects are also regarded as less probable than $C_N$,[19,20] we recently identified di- and tri-carbon defects.[21,22] The proof of the defects' nature was based on an analysis of local vibrational modes (LVMs) observed in Raman scattering or infrared (IR) absorption spectra of AlN crystals containing carbon in natural isotopic and $^{13}C$ isotope enriched composition. Furthermore, we reported the discovery of IR absorption lines at 1678 cm$^{-1}$ and 1718 cm$^{-1}$ in GaN doped with naturally abundant carbon,[23] and proved the vibration nature of these lines by the isotope shift observed in a sample enriched to nearly 100% with the isotope $^{13}$C.[7] The underlying carbon-related defects start to form for carbon concentrations [C] > 5×10$^{17}$ cm$^{-3}$ and presumably dominate over other carbon defect species for [C] > 10$^{19}$ cm$^{-3}$. Since the antisymmetric stretching vibration modes of the tri-carbon defect in AlN at 1769 cm$^{-1}$ (Ref. 22) and of the $C_3^-$ carbon cluster anion in an argon matrix at 1721.8 cm$^{-1}$ (Ref. 24) possess similar wavenumbers as the IR absorption lines in GaN, it is reasonable to assume that they likewise arise from tri-carbon defects.[23] However, to basically understand the electrical compensation in GaN:C, unequivocal evidence for the substantial formation of electrically active tri-carbon defects for moderate to heavy carbon doping of GaN remains to be given.

In the present paper, we show that the infrared absorption lines at 1678 cm$^{-1}$ and 1718 cm$^{-1}$ in carbon doped GaN (GaN:C) are due to LVMs originating from defects consisting of three carbon atoms each. The proof is based on the analysis of the isotope effect on the LVMs observed in GaN crystals containing carbon of distinct $^{12}$C/$^{13}$C isotope ratios. Combined with the polarization dependence of the LVM absorption it is shown, that the tri-carbon defects form a triatomic molecule-like structure in two crystallographically different configurations. The LVMs disappear under additional illumination with photon energies just below the bandgap energy. This is interpreted as a change of the charge state of the tri-carbon defects and proves their contribution to optical absorption as well as to electrical charge balance.

II. EXPERIMENTAL

Thick, +$c$-plane oriented GaN layers, carbon doped and undoped ones, were grown by hydride vapor phase epitaxy (HVPE) on 2″ (0001) GaN/sapphire templates. The growth and doping procedure is described in detail in Ref. 7. For the carbon doping we used the following sources: liquid pentane (Dockweiler Chemicals, electronic grade) or gaseous butane (Sigma Aldrich, Butane 12C4 (Gas) 99%) containing carbon in the natural isotopic composition. Both show the same carbon doping efficiency. Additionally, a butane source isotopically enriched with $^{13}$C to 99% (Sigma Aldrich, Butane 13C4 (Gas) 99%) was used to achieve a doping with 99% and 50% of the isotope $^{13}$C. All layers separated from the sapphire substrates spontaneously after cool-down due to the large difference of the thermal expansion coefficients. $C$-plane oriented samples of about 10×5×0.5–1 mm$^3$ were diced and both $c$-faces were polished. From two of samples (s2 and c1213, see Table I) cross-sectional $m$-plane stripes of about 5×1×0.5 mm$^3$ were cut and polished. Concerning the crystalline perfection of the samples, we refer to the detailed investigation reported recently.[7] In-plane and out-of-plane strain of the $c$-plane GaN layers are negligible (accuracy of the measurement of the lattice constants by X-ray diffraction: ±0.03% and ±0.01%, respectively), and the full width at half maximum of X-ray diffraction rocking curves at the symmetric 002 reflection and the skew-symmetric 302 reflection varies between 80 and 120 arcsec. Since the results of both these measurements are essentially independent of whether the layers are carbon doped or not, carbon doping does not impact the crystalline quality.

Carbon, oxygen, silicon, and hydrogen concentrations ([C], [O], [Si], and [H]) of the GaN layers were determined by secondary ion mass spectrometry (SIMS, performed by RTG Mikroanalyse GmbH Berlin). In the undoped reference sample [C], [O], and [Si] are below the respective SIMS detection limits: [C] < 2.4×10$^{16}$ cm$^{-3}$, [O] < 2×10$^{16}$ cm$^{-3}$, and [Si] < 7×10$^{15}$ cm$^{-3}$. In the pentane doped samples [O] and [Si] are slightly raised above the detection limit to 3.5×10$^{16}$ cm$^{-3}$ and

$1.3\times10^{16}$ cm$^{-3}$, respectively, while in the samples doped with isotopically enriched butane, [O] attains $3.5\times10^{17}$ cm$^{-3}$. [H] is at about $1\times10^{17}$ cm$^{-3}$ for undoped GaN layers and does not exceed $6\times10^{17}$ cm$^{-3}$ for the carbon doping levels of the samples of the present investigation. The total carbon content [C]$_{tot}$ = [$^{12}$C] + [$^{13}$C] and the respective $^{13}$C percentage of each sample are specified in Table I. Comparing [C]$_{tot}$ with the residual impurity concentrations mentioned above, [O] and [Si] attain at most one tenth of [C]$_{tot}$, while [H] could reach one third. Hence, compensation or passivation of carbon acceptors by these impurities are expected to play no major role here.

Fourier-transform infrared (FTIR) absorption measurements were performed in the mid-infrared spectral range on a Bruker Vertex 80v spectrometer equipped with a globar source, a potassium bromide beam splitter, and a liquid nitrogen cooled mercury cadmium telluride detector. For measurements below room temperature, a liquid-helium-flow cryostat (Oxford OptistatCF) with zinc selenide windows was used. The spectra were recorded with a resolution of 0.5 cm$^{-1}$ at room temperature and 0.25 cm$^{-1}$ at 10 K. For polarization dependent measurements, a holographic wire grid polarizer on a KRS-5 substrate was used. The reference sample (ref) with carbon level below the SIMS detection limit was used for subtraction of the background due to intrinsic absorption bands. Additional FTIR measurements were carried out under below-bandgap illumination. The exciting light of wavelength 385 nm (3.2 eV) was guided by an optical fiber from a power light-emitting-diode source (Omicron LedHUB) into the sample chamber of the FTIR spectrometer using a vacuum-tight feed-through. The fiber output was positioned in such a way that it was outside of the FTIR sample beam and that the LED light was incident under ~60 degrees to the sample surface. Nominal LED powers between 7.5 mW and 250 mW were used in continuous mode.

## III. RESULTS

FTIR absorption spectra of undoped GaN and GaN:C $c$-plane samples are shown in Fig. 1 in the typical range of carbon cluster vibrations.[24] The strong absorption due to intrinsic one-phonon and two-phonon-combination vibrations in the range below 1400 cm$^{-1}$ limits the useful spectral range of IR absorption to higher wavenumbers since relatively thick samples have been measured for improved detectability of extrinsic vibration modes. The measurements were carried out with the incident light beam perpendicular to the $c$-facet. Thus, the angle $\varphi$ between the electric field vector $\boldsymbol{E}$ of the light wave and the crystal's $c$-axis is equal to 90° (polarization $\boldsymbol{E} \perp c$) irrespective of whether the light wave is linearly or circularly polarized. The two absorption lines labeled 12A and 12B in Fig. 1, suggested earlier to be due to carbon-related LVMs,[23] increase in intensity with carbon concentration.

LVMs 12A and 12B have similar height at $\boldsymbol{E} \perp c$ (Fig. 1), but polarization dependent measurements through the $m$-facet of sample s2 at room temperature (Fig. 2 (a, c)) exhibit pronounced differences for both peaks. Consistent with the measurement presented in Fig. 1, the spectrum of the $m$-plane sample for $\boldsymbol{E} \perp c$ in Fig. 2(a) shows both peaks with comparable height too, whereas in the spectrum for $\boldsymbol{E} \parallel c$ peak 12A becomes much higher and peak 12B disappears. The complete dependence of the LVM peak absorbance on the polarization angle $\varphi$ in Fig. 2(c) reveals that peaks 12A and 12B attain a non-zero minimum and a maximum, respectively, at $\varphi = 90°$ ($\boldsymbol{E} \perp c$), while for $\varphi = 0°$ ($\boldsymbol{E} \parallel c$) peak 12A reaches a maximum and peak 12B becomes zero.

Mode 12B occurs at 1679 cm$^{-1}$ at 297 K, shifts significantly (to 1673.9 cm$^{-1}$) when cooled to 10 K, and becomes much narrower in contrast to mode 12A which maintains its position (at 1718 cm$^{-1}$) and width at both temperatures (Fig. 2 (b)).

Figure 3 shows IR absorption spectra of samples c12, c13, and c1213 recorded at low temperature (10 K). Fig. 3 (a) displays LVMs 12A and 12B in sample c12 (~99% $^{12}$C) together with two new LVMs (13A and 13B) in sample c13 (~99% $^{13}$C). 13A and 13B appearing at 1651.3 and 1610.5 cm$^{-1}$ in the spectrum have identical shape as 12A and 12B. Ten additional peaks are detected for sample c1213 with the mixed isotope composition. In the spectra captured at $E \perp c$ (Fig. 3 (b)), the peaks 1 and 6, corresponding to modes 12B and 13B, form two triplets of intensity ratio 1:2:1 with the new peaks (2 – 5) (stick spectra under the curve, Table II). In the spectra for $E \parallel c$ shown in Fig. 3 (c), peak 1 (12A), peak 8 (13A) and six new peaks 2 – 7 (Table II) are grouped to two quadruplets of 1:1:1:1 intensity marked as the red stick spectra, while the modes dominating in Fig. 3(b) are effectively suppressed for this polarization. We note that although the peaks from 5 to 8 seem higher than 1 - 4, the total area of the peaks 1 – 4 is equal to that of the peaks 5-8.

Finally, the intensity of peaks 12A and 12B decreases under below-bandgap excitation at 385 nm (3.22 eV) as shown in Fig. 4. The signals completely disappear for high intensity excitation.

## IV. DISCUSSION

Shift (Fig. 3 (a)) and splitting (Fig. 3 (b, c)) of the infrared absorption lines observed in the spectral range between 1600 cm$^{-1}$ and 1750 cm$^{-1}$ due to the $^{13}$C isotope enriched doping of samples c13 and c1213 prove the vibrational nature of these absorption lines. Key for this interpretation is the application of the model of a harmonic spring oscillator to estimate the frequency change of a molecule-like defect upon the replacement of isotopes within the defect structure, i.e. by forming different isotopomeric defects. Assuming an unchanged force constant, the ratio of the vibration frequencies $v$ of two isotopomers $i$ and $j$ is given by $v_i/v_j = \sqrt{m_j/m_i}$, where $m_i$ and $m_j$ are the respective reduced masses. If the defect exclusively consists of $n$ carbon atoms and its vibration only negligibly interacts with the surrounding host lattice atoms, then in the case of complete replacement of $n$ $^{12}$C isotopes by $n$ $^{13}$C isotopes a frequency ratio of $\sqrt{13/12} \approx 1.041$ is expected, i.e. the wavenumber of the $^{13}$C isotopomer LVM is accordingly lowered. This theoretical maximum value has been closely observed e.g. for the frequency ratio of the vibration modes of elementary anionic tri-carbon clusters in isotope exchange experiments.[24] Foreign atoms if involved into the defect structure which are not substituted in the isotope replacement may only decrease the frequency ratio. For complete $^{12}$C/$^{13}$C isotope exchange between samples c12 and c13, the wave numbers of the LVM pairs 12A, 13A and 12B, 13B, respectively (Fig. 3 (a)), give nearly the same ratio 1717.8/1651.5 ≈ 1673.9/1610.5 = 1.0395±0.005. This ratio is very close to the maximum possible one suggesting that the LVMs originate from a molecule-like defect formed by carbon atoms only. The slightly lowered value of the ratio may be ascribed to the influence of the surrounding host atoms.

The question how many carbon atoms $n$ form the defect can be answered by analyzing the line splitting of the LVMs in sample c1213 doped with the $^{12}$C/$^{13}$C isotope mixture (Fig. 3 (b, c)). In the general case that $n$ inequivalent lattice sites within the defect structure are available for the distribution of the two carbon isotopes, one obtains $2^n$ combinations of differing isotope arrangements. Hence, the corresponding isotopomeric defects generally possess $2^n$ different frequencies per LVM. Since in the special case of sample c1213 both isotopes are equally abundant (each at 50% of the total carbon concentration), one expects that each of the isotopomeric defects forms with equal probability and that the individual LVMs have the same intensity. If some of the $n$ lattice sites are equivalent, then not all of the $2^n$ combinations differ in their isotope arrangement and hence, some of the individual LVMs are degenerate and possess correspondingly higher intensity.

The splitting of LVMs 12A/13A into eight lines of equal intensity (integrated peak area) in the FTIR spectrum of sample c1213 (Fig. 3 (c), Table III) suggests that $n = 3$ carbon atoms occupy crystallographically inequivalent sites (XYZ, Fig. 5) in the defect structure leading to $2^3 = 8$ isotopomeric defects with individual LVM frequencies. The six-fold line splitting of modes 12B/13B (Fig. 3(b)) can also be explained by a defect containing three carbon atoms. But now the carbon atoms occupy two equivalent sites (Y) and a third unique site (X) in between (XY$_2$, Fig. 5). The possible isotope distributions also give rise to eight isotopomeric defects but two of them are identical in two cases, so that two of the individual LVMs (2 and 5 in Fig. 3(b)) are doubly degenerate. The observed splitting pattern of two triplets each with 1:2:1 intensity ratio (Fig. 3(b), Table II) is fully consistent with this defect model. Conclusion from the analysis so far is, that the LVMs at 1679 cm$^{-1}$ and 1718 cm$^{-1}$ dominating in crystals doped with carbon of natural isotopic composition originate from two defects each consisting of three carbon atoms forming a molecule-like cluster.

The polarization dependence of the LVM absorption (Fig. 2(c)) gives information on the orientation of both tri-carbon defect configurations with respect to the crystal coordinate system. More precisely, one can determine the orientation of the oscillating electric dipole moment $\mathbf{d}$ belonging to the actual normal mode. For this purpose, we derive in the appendix a formula describing the vibrational absorption $A$ as a function of both the angle $\alpha$ between $\mathbf{d}$ and the GaN crystal's $c$-axis and the angle $\varphi$ between $\mathbf{E}$ and $c$

$$A(\varphi, \alpha) = A_0(\sin^2\alpha + (2 - 3\sin^2\alpha) \cdot \cos^2\varphi) \tag{1}$$

Using Eq. (1) for the calculation of the polarization dependence shown in Fig. 2(c), the tilt angle $\alpha$ can be determined by best fitting of the calculated curve to the experimental data. The polarization dependence of the LVM 12B is best described by $\alpha = 90°$ meaning that the direction of the oscillating electric dipole moment is parallel to the $c$-plane. This in particular includes the directions [01.0], [10.0], and [11.0], which are parallel to the connection lines of each two nearest N (or Ga) atoms in the basal plane of the N (or Ga) sublattice (cf. Fig. 5(a)). For the LVM 12A, $\alpha$ is found to be 33° ± 1°, a value that is very close to the angle 35.38° (calculated using the VESTA program[25] and lattice parameters relevant for GaN[26]) between the $c$-axis and the directions of the lines connecting two nearest N (or Ga) atoms located in two sublattice $c$-planes shifted by $c/2$ (corresponding to the directions [42.-3], [42.3], [24.-3], [24.3], [2-2.3], and [-22.3]). In order to relate the direction of the oscillating dipole moment to the geometry of the tri-carbon defects, it is necessary to know the type of normal mode observed. Since the wave numbers of the LVMs 12A and 12B are in the range around 1700 cm$^{-1}$, as is the case for the antisymmetric stretching mode of the tri-carbon defect in AlN and the $C_3^-$ carbon cluster anion,[22,24] we suggest that the LVMs arise from antisymmetric stretching modes and not from the other two possible normal vibrations of triatomic molecules, the symmetric stretching mode and the bending mode. Furthermore, these latter two modes are generally expected to possess lower vibrational energies and may be shifted to wavenumbers below 1400 cm$^{-1}$ which is out of the detection window of our experiments. The oscillating dipole moment of the antisymmetric stretching mode is aligned along the line between the two end atoms of the triatomic molecule-like defects in the symmetrical XY$_2$ case (C$_{2v}$ molecule symmetry). This alignment is not perfect in the asymmetrical XYZ case (C$_s$ molecule symmetry) where the two directions may be tilted to each other within the molecule's symmetry plane. However, this tilt is expected to be small in the present case of a molecule-like defect consisting of three atoms of the same element. Thus, it is most reasonable to assume that the two end atoms of the tri-carbon defects substitute two nearest N (or Ga) atoms within the same $c$-plane in the case of the defect configuration associated with LVM 12B, while in the case of the defect configuration associated with LVM 12A they substitute two nearest N (or Ga) atoms located in two $c$-planes which are shifted by $c/2$. The former defect configuration we call the basal tri-carbon defect, the latter one the axial tri-carbon defect. Both defect configurations are depicted in Fig. 5.

In the case of the antisymmetric stretching mode of a triatomic molecule, the isotope effect also provides information on the angle between both bonds of the molecule. Herzberg derived a formula relating the corresponding frequencies of isotope exchanged symmetrical XY$_2$ molecules in dependence on atomic masses and bond angle:[27]

$$\left(\frac{v_i}{v_j}\right)^2 = \frac{m_j^X m_j^Y \left(m_i^X + 2m_i^Y \sin^2\beta\right)}{m_i^X m_i^Y \left(m_j^X + 2m_j^Y \sin^2\beta\right)} \qquad (2)$$

Here, $2\beta$ is the angle between two X-Y bonds (see the illustration in Fig. 5(b)), the subscripts $i$ and $j$ correspond to the isotopomers before and after the isotope substitution, $v_{i,j}$ are the individual frequencies (or wavenumbers) of the antisymmetric stretching mode, $m_{i,j}^{X,Y}$ are the atomic masses (equal to 12 u or 13 u for the two carbon isotopes) of the central (X) or the end (Y) carbon atoms. We apply Eq. (2) to wavenumbers of the antisymmetric stretching mode of the isotopomers of the basal tri-carbon defect ($C_{2v}$ molecule symmetry), which are summarized in Table II. Taking one of the wavenumbers as reference, e.g. $v_1$ from Table II, the wavenumbers for such isotope replacements, which conserve the $C_{2v}$ molecule symmetry, can be calculated: replacement of the central atom ($v_4$), of both Y atoms ($v_3$), and of all three atoms ($v_6$). Thereby, the bonding angle $2\beta$ is treated as fitting parameter. The best fit to the experimental line positions (mismatch below 1.6 cm$^{-1}$) is obtained for $2\beta = 134\pm8°$. For the two missing modes ($v_2, v_5$), when the two Y sites are occupied by different isotopes, the difference in isotope properties does not break the symmetry significantly. Therefore, one expects their frequency positions half the way between the modes for the substitution of both Y atoms: $v_2 \approx (v_1 + v_3)/2$ and $v_5 \approx (v_4 + v_6)/2$, also in good agreement with experiment. When Eq. (2) is used for the calculation of the individual LVM positions of the axial tri-carbon defect of lower molecule symmetry $C_s$ (XYZ, Table III), the experimental LVMs 1, 4, 5, and 8 are in best agreement with the model for $2\beta = 154°$. The frequencies for the replacement of only one end atom are no longer degenerated; however, the experimental peaks 2, 3 and 6, 7, respectively, stay equidistant from the center of the quadruplets (green dash-dot lines in Fig. 3 (c)) and still $(v_2 + v_3)/2 \approx (v_1 + v_4)/2$ and $(v_6 + v_7)/2 \approx (v_5 + v_8)/2$. Since Eq. (2) is not strictly valid for the case of the asymmetric axial configuration of the tri-carbon defect, we allow for a mismatch of 2 cm$^{-1}$ between the experimental and calculated line positions to realistically estimate a lower limit of the bond angle at $2\beta > 140°$ (instead of $2\beta = 154°$).

So far, we have focused on the isotope effect observed in sample c1213 containing $^{12}$C and $^{13}$C isotopes in equal proportion. Therefore, all individual LVMs of the isotopomeric tri-carbon defects are clearly visible in this sample with equal (or doubled, if degenerated) intensity. In the samples c12, s1, s2, and c13, where one isotope dominates with about 99% abundance, the mixed isotopomers have strongly reduced formation probability. Among them, isotopomers formed by two major isotopes and a minor one should have the highest probability of about 1% (0.99×0.99×0.01 = 0.0098). In fact, we find in these samples absorption lines of the expected low intensity which coincide with some of the LVM positions found in sample c1213. The corresponding wavenumbers are listed in Tables II and III, and the LVM peaks are marked by asterisks in Figs. 2 and 3. The observation of these LVM peaks of minor intensity additionally supports our assignment of the various isotopomers to their individual LVMs.

Concerning possible models for the arrangement of the three C atoms, we have already concluded from the polarization dependence that the end C atoms of both tri-carbon defect configurations most likely occupy nearest-neighbor N (or Ga) sites of the N (Ga) sublattice without large displacements. With view on the wurtzite lattice of GaN (see Fig. 5(a)), it seems reasonable to position the third, middle C atom on the site of the complementary host atom, i.e. on Ga (or N) site. In this case, however, one would expect a bonding angle equal to that of GaN of about 109°. This is not in agreement with the bonding angles we obtain from the analysis of the individual LVM positions using Eq. (2) (calculations assuming $2\beta =109°$ lead to a mismatch of 5 and 7 cm$^{-1}$ for the basal and axial defect, respectively). The calculated $2\beta$ angles are much larger ($134 \pm 8°$ for the basal configuration and $> 140°$ for the axial configuration, Fig. 5(b)) and suggest a considerable displacement of the middle C atom from the substitutional site to the line connecting the outer C atoms. The corresponding local lattice relaxation might be driven by the shorter bond length between C atoms in comparison to the bond length between Ga and N atoms. Supposing that C atoms prefer to occupy N over Ga sites, we suggest the tri-carbon defects $C_N$ – $C_{Ga\text{-}off}$ – $C_N$ ($C_{Ga\text{-}off}$ means off-center Ga site) in two crystallographically inequivalent configurations, the basal and the axial one, to be of major abundance and responsible for the LVMs 12A (1718 cm$^{-1}$) and 12B (1678 cm$^{-1}$), respectively. The counterparts of the former tri-carbon defects, having the structural formula $C_{Ga}$ – $C_{N\text{-}off}$ – $C_{Ga}$ ($C_{N\text{-}off}$ means off-center N site), are expected to form with lower probability, but to vibrate at similar frequencies as LVMs 12A and 12B. Therefore, we tentatively attribute the weak satellite peaks of 12A and 12B found at 1736.6 cm$^{-1}$ and 1681.6 cm$^{-1}$ (Figs. 2 (b), 3 (a)) as well as their isotope $^{13}$C related twins at 1680.9 cm$^{-1}$ and 1616.8 cm$^{-1}$ to $C_{Ga}$ – $C_{N\text{-}off}$ – $C_{Ga}$ defects.

The temperature dependence of shape and position of the LVMs 12A and 12B demonstrates different interaction of the basal and axial tri-carbon defects with the lattice vibrations. LVM 12B shifts its position to lower wavenumbers and appreciably narrows with decreasing temperature while in contrast LVM 12A hardly shows any changes (Fig. 2). Analogous behavior was observed, for example, in Raman investigations of Mg-H complexes in InN:Mg and in IR absorption of Se-related complexes in AlSb:Se,H,D,[28,29] which the authors explained by resonant interactions between the defect vibration and optical phonons. Since the basal tri-carbon defect configuration is symmetric ($C_{2v}$ molecule symmetry), its antisymmetric stretching mode might couple more strongly with the phonon system than that of the asymmetric axial configuration.

Finally, the performed excitation experiment (Fig. 4) allows to claim that the tri-carbon defects are electrically active. The sub-bandgap excitation at 385 nm (3.22 eV) leads to charge carrier redistribution at the defects. Consequently, the tri-carbon defects change the charge state either by direct photoionization or by capturing carriers generated by the optical excitation of other defects (e.g. $C_N$). In the altered charge state, the antisymmetric stretching modes of the tri-carbon defects may be unobservable due to local changes of the charge distribution between the carbon atoms strongly reducing the intensity of that modes, shifting the frequency or leading to line broadening. Consequently, the tri-carbon defects possess at least one charge state transition level within the bandgap and contribute to optical absorption as well as to the electrical charge balance. It follows that the LVM intensity depends on the charge state and hence, on the position of Fermi or quasi-Fermi levels.

## V. SUMMARY

Based on the direct method of analyzing the isotope effect, the LVMs at 1678 cm$^{-1}$ (12B) and 1718 cm$^{-1}$ (12A) in GaN:C are unambiguously associated with tri-carbon defects in basal and axial configurations. The basal configuration with six-line isotope splitting shows $C_{2v}$ molecule symmetry due to the two end atoms on equivalent positions. The axial configuration with eight-line splitting possesses lower molecule symmetry $C_s$. The exact spectral positions of the individual LVMs arising from the isotope effect can be well explained within the model of the harmonic oscillator applied to the antisymmetric stretching modes of bent triatomic molecule-like defects. This analysis also allows to estimate the angle $2\beta$ between the C-C bonds, equal to 134 ± 8° for the basal and over 140° for the axial configuration, and thus is much larger than the corresponding bond angle of ~109° in the GaN crystal lattice, which is reasonable for shorter carbon-carbon bonds. Hence, the lattice itself is locally distorted significantly and at least one of three carbon atoms is shifted from the normal lattice site. However, the most probable positioning of the two carbon end atoms is inferred from the polarization dependence of the LVMs 12A and 12B to be a substitution of two nearest N or Ga host atoms. Therefore, we suggest that the preferred site occupation of the tri-carbon defects is $C_N - C_{Ga\text{-off}} - C_N$ in the two crystallographically inequivalent configurations associated with LVMs 12A and 12B, while the complementary site occupation $C_{Ga} - C_{N\text{-off}} - C_{Ga}$ may be tentatively associated with minor LVMs at 1736.6 and 1681.6 cm$^{-1}$ (at 10 K) appearing only for the highest carbon concentrations. LVMs 12A and 12B change intensity under light excitation at 385 nm, which proves that the defect charge state changes. Taking into account the dipole sum rule for an order-of-magnitude estimation of the defect density given in our previous paper,[23] we conclude, that the described tri-carbon complexes possibly determine the optical and electrical properties of the material in samples with high carbon concentrations.

ACKNOWLEDGEMENT

One of the authors (IG) was supported by Deutsche Forschungsgemeinschaft (DFG) under project contract BI 781/11.

APPENDIX

A supporting sketch is shown in Fig. 5(b). In the dipole approximation, the absorption $A(\varphi, \alpha)$ as a function of the angles $\alpha$ (between the oscillating dipole and wurtzite symmetry axis $c \parallel z$) and $\varphi$ (between $\boldsymbol{E}$ and $c$) is proportional to the square of the scalar product of electric field vector $\boldsymbol{E}$ and $\boldsymbol{d}$: $A(\varphi, \alpha) \sim (\boldsymbol{E} \cdot \boldsymbol{d})^2$. Thus, the second power in this dependence allows simplifying the natural GaN six-fold screw rotation axis to C3 axis (parallel to $z$) and express $\boldsymbol{d}$ by not six but by three vectors $\boldsymbol{t}, \boldsymbol{r}$, and $\boldsymbol{f}$ corresponding to three possible equivalent configurations of this defect satisfying the axis symmetry requirements. We also introduce distance Z – Y: $|\boldsymbol{t}| = |\boldsymbol{r}| = |\boldsymbol{f}| = d$. Thus, the absorption $A$ is proportional (1) to a sum of squares of projections of vectors $\boldsymbol{t}, \boldsymbol{r}$, and $\boldsymbol{f}$ to $\boldsymbol{E}$. The vectors' coordinates $x$ or $y$ are shown in units of $l$ when $z$ is in the units of lattice parameter $c$. Concerning $(|\boldsymbol{E}|_t)^2 = (t_x \cos\theta + t_y \cos\tau + t_z \cos\varphi)^2$ and the analogues for $|\boldsymbol{E}|_r$ and $|\boldsymbol{E}|_f$, the conversion (2) is gotten (the projection definition is applied with angles between $\boldsymbol{E}$ and axes $\boldsymbol{x}, \boldsymbol{y}, \boldsymbol{z}$ are $\theta, \tau, \varphi$ when $\varphi = 90° - \theta$; $\tau = 90°$ if $\boldsymbol{E}$ turns around $\boldsymbol{y}$ in the plane $xz$). Then, placing the coordinates (i.e. $t_x = -\frac{l}{2}, t_y = -\frac{\sqrt{3}\, l}{2}, t_z = \frac{c}{2}$, etc.), and considering the natural relations $l = d \cdot \sin\alpha, c = 2d \cdot \cos\alpha$, then, opening the squares and seeking the simplest form (4), we obtain the explicit dependence $A(\varphi, \alpha)$ of the absorbance (5) after the final trigonometric simplifications and hiding of all auxiliary constants into normalization constant $A_0 = 3/2 \cdot d^2$:

$$A(\varphi, \alpha) \stackrel{(1)}{\sim} \left(|\vec{E}|_t\right)^2 + \left(|\vec{E}|_r\right)^2 + \left(|\vec{E}|_f\right)^2 \stackrel{(2)}{=} (t_x \sin\varphi + t_y \cos 90° + t_z \cos\varphi)^2 + \cdots \stackrel{(3)}{=}$$

$$\stackrel{(3)}{=} \frac{3}{2} l^2 \sin^2\varphi + \frac{3}{4} c^2 \cos^2\varphi \stackrel{(4)}{=} \frac{3}{2} d^2 (\sin^2\alpha \cdot \sin^2\varphi + \cos^2\alpha \cdot \cos^2\varphi) \stackrel{(5)}{=}$$

$$A(\varphi, \alpha) = A_0 (\sin^2\alpha + (2 - 3\sin^2\alpha) \cdot \cos^2\varphi)$$

TABLE I. Total carbon concentration and isotope content of the investigated GaN samples as determined by SIMS.

| Sample | $[^{12}C] + [^{13}C]$ (cm$^{-3}$) | $[^{13}C] / ([^{12}C] + [^{13}C])$ (%) |
|---|---|---|
| c12 | $5.8 \times 10^{18}$ | ~1% |
| c13 | $2.5 \times 10^{18}$ | ~99% |
| c1213 | $5.1 \times 10^{18}$ | ~50% |
| s1 | $1.9 \times 10^{18}$ | ~1% |
| s2 | $1.4 \times 10^{19}$ | ~1% |
| ref | $< 2.4 \times 10^{16}$ | ~1% |

TABLE II. Individual LVM wavenumbers of different isotopomers for the basal tri-carbon defect with molecule symmetry $C_{2v}$

| | $\nu_i$, cm$^{-1}$ | | | | |
|---|---|---|---|---|---|
| $i$ | c12 T=10 (297) K | c13 T=10 K | c1213 T=10 K | Calculations $2\beta= 134.8°$ | Isotopomer |
| 1 (12B) | 1673.9 (1679) | - | 1673.9 | 1675.5 | $^{12}C$-$^{12}C$-$^{12}C$ |
| 2 | | - | 1662.3 | 1663.3 | $^{13}C$-$^{12}C$-$^{12}C$ / $^{12}C$-$^{12}C$-$^{13}C$ |
| 3 | 1651.5 | - | 1651.5 | 1651.1 | $^{13}C$-$^{12}C$-$^{13}C$ |
| 4 | 1634.4 | - | 1634.1 | 1633.9 | $^{12}C$-$^{13}C$-$^{12}C$ |
| 5 | - | 1622.5 | 1622.5 | 1621.5 | $^{13}C$-$^{13}C$-$^{12}C$ / $^{12}C$-$^{13}C$-$^{13}C$ |
| 6 (13B) | - | 1610.5 | 1610.5 | 1609.1 | $^{13}C$-$^{13}C$-$^{13}C$ |

Table III. Individual LVM wavenumbers of different isotopomers for the axial tri-carbon defect with molecule symmetry $C_s$

| $i$ | $\nu_i$, cm$^{-1}$ | | | | Isotopomer |
|---|---|---|---|---|---|
| | c12 T=10 (297) K | c13 T=10 (297) K | c1213 T=10K | Calculations $2\beta= 154°$ | |
| 1 (12A) | 1717.8 (1718) | - | 1717.8 | 1718.5 | $^{12}C$-$^{12}C$-$^{12}C$ |
| 2 | - | - | 1714.4 | 1716.9$^\dagger$ | $^{12}C$-$^{12}C$-$^{13}C$ |
| 3 | 1699.2 | - | 1699.0 | 1699.9$^\dagger$ | $^{13}C$-$^{12}C$-$^{12}C$ |
| 4 | - | - | 1694.6 | 1695.3 | $^{13}C$-$^{12}C$-$^{13}C$ |
| 5 | 1675.8 (1676) | - | 1675.8 | 1674.3 | $^{12}C$-$^{13}C$-$^{12}C$ |
| 6 | - | - | 1672.4 | 1670.4$^\dagger$ | $^{12}C$-$^{13}C$-$^{13}C$ |
| 7 | - | - | 1656.5 | 1654.9$^\dagger$ | $^{13}C$-$^{13}C$-$^{12}C$ |
| 8 (13A) | - | 1651.5 (1652) | 1651.7 | 1651.0 | $^{13}C$-$^{13}C$-$^{13}C$ |

$^\dagger$estimated from the principal equidistance of these modes from the position of the degenerated mode of a symmetrical defect 1708.4 and 1662.6 cm$^{-1}$ (green dash-dot lines in Fig. 3(c))

FIGURE CAPTIONS

FIG. 1. FTIR room-temperature spectra of GaN *c*-plane samples doped with carbon of natural isotopic composition in comparison to the spectrum of the undoped reference sample. The strongly increasing absorption towards lower wavenumbers is due to two-phonon combinations of GaN intrinsic vibrations. The local vibrational modes labeled 12A and 12B increase in intensity with the carbon concentration and are absent in the reference sample.

FIG. 2. FTIR spectra of the LVMs 12A and 12B measured through the *m*-facet of sample s2 (a) at room temperature and (b) at T = 10 K in the extreme cases of polarization ***E*** ∥ *c* ($\varphi = 0°$) and ***E*** ⊥ *c* ($\varphi = 90°$). (c) Complete polarization dependence of LVMs 12A and 12B obtained at 297 K. Spectral positions marked with asterisk (*) coincide with LVMs observed in sample c1213 (see Tables II and III as well as Fig. 3(b,c)).

FIG. 3. Low-temperature FTIR spectra (normalized to the respective LVM intensity maximum), illustrating the isotope effect (a) in the samples c12 and c13 as well as (b) and (c) in the sample c1213 for the two linear polarizations of the incident light ***E*** ⊥ *c* and ***E*** ∥ *c*, respectively. The splitting (b) to a pair of triplets and (c) to a pair of quadruplets (clearly seen here by using differently polarized light) is consistent with tri-carbon defects possessing $C_{2v}$ and $C_s$ molecule symmetry, respectively. Areas of the peak groups 1-4 and 5-8 in (c) are equivalent, i.e. all absorption lines have equal intensity. The minor isotope abundance of $^{13}C$ and $^{12}C$ (~1%) in samples c12 and c13, respectively, causes LVMs of weak intensity coinciding with the peaks marked by an asterisk (*) (see also Fig. 2).

FIG. 4. FTIR spectra of sample s1 without background subtraction under additional light excitation at 385 nm. The LVMs 12A and 12B disappear partly for low-intensity and completely for high-intensity excitation.

Fig. 5. (a) Illustrations of the tri-carbon $C_N$-$C_{Ga}$-$C_N$ defect (ideal case with the central C atom on regular substitutional site) in two crystallographically inequivalent configurations, axial XYZ (A) and basal XY$_2$ (B) in the GaN crystal lattice (visualized using the VESTA software[25]). (b) Illustrations of the tri-carbon XY$_2$ or XYZ defects with experimentally found angles $2\beta \gg 109°$ in the simplest assumption of substitutional end atoms Y, Z and displaced central atoms X. The green ball marks the regular substitutional lattice position. The direction of the oscillating dipole moment $\boldsymbol{d}$ in the case of the antisymmetric stretching mode of three crystallographically equivalent XYZ defect configurations is given by the vectors $\boldsymbol{t}, \boldsymbol{r}, \boldsymbol{f}$ (see appendix).

FIGURE 1

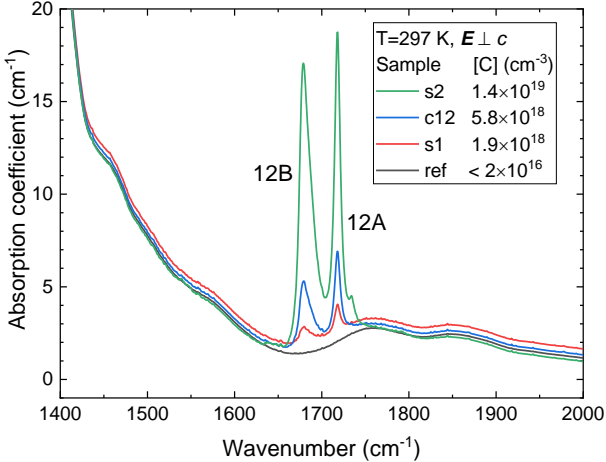

FIGURE 2

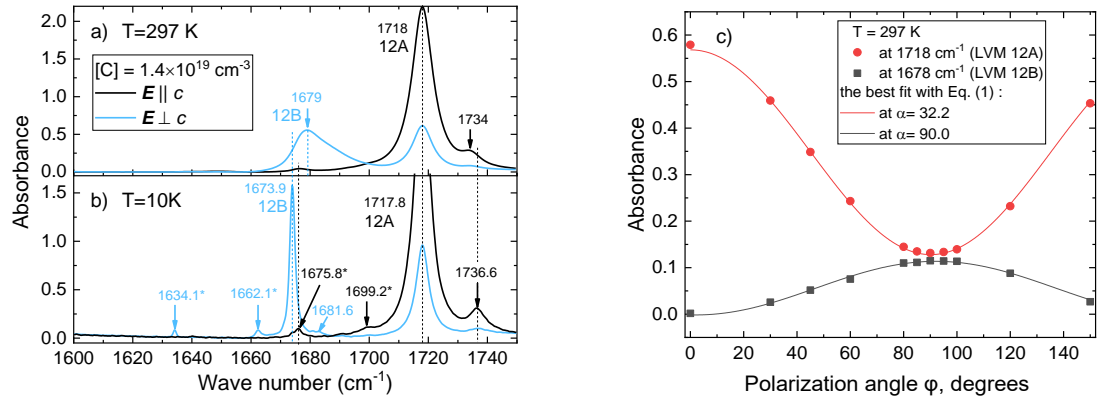

FIGURE 3

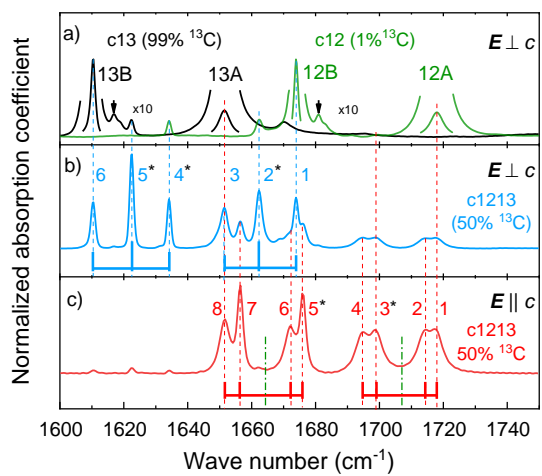

FIGURE 4

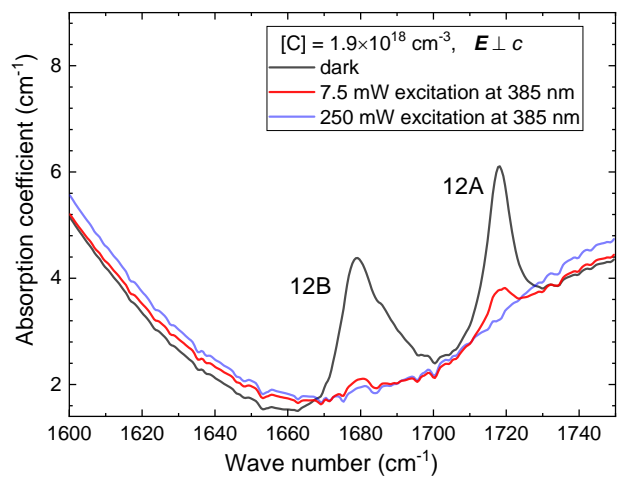

FIGURE 5

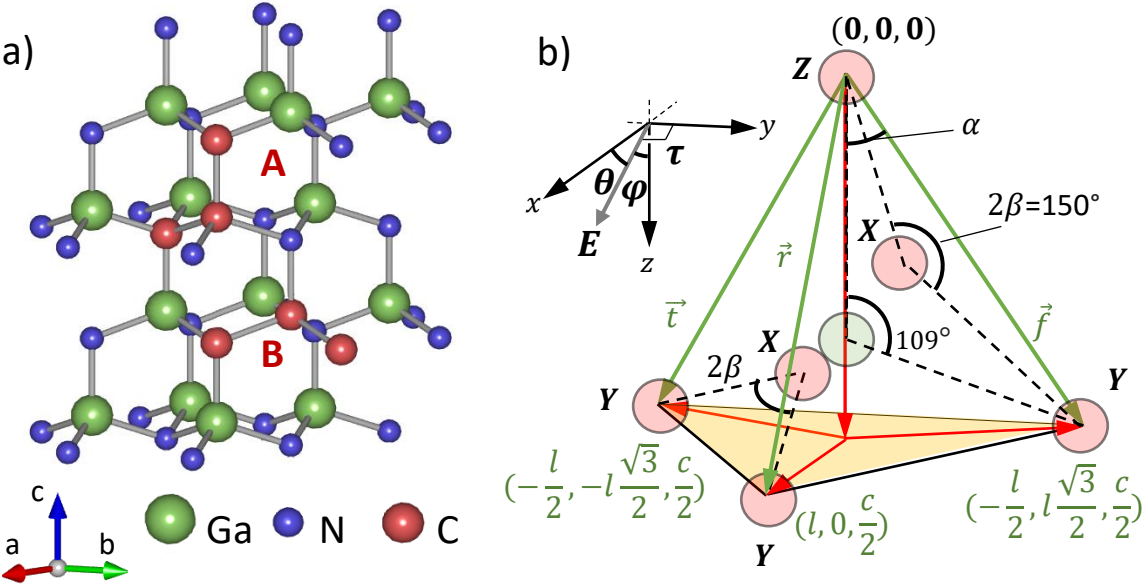